\documentclass{pnastwo}

\usepackage{pnastwof}
\usepackage{graphicx}
\usepackage{amssymb,amsfonts,amsmath}
\usepackage{color}
\usepackage{soul}

\newcommand{\alopt}{\alpha_{\text{opt}}}
\newcommand{\vc}[1]{\mathbf{#1}}
\newcommand{\vcv}{\vc{v}}
\newcommand{\disrate}{\mathit{D}}
\newcommand{\Rent}{R^{\text{ent}}}
\newcommand{\Rcone}{R^{\text{cone}}}
\newcommand{\Rentcyl}{R^{\text{ent,cyl}}}

\newcommand{\comm}[1]{{#1}}
\newcommand{\red}[1]{{#1}}
\newcommand{\blue}[1]{{#1}}
\newcommand{\lb}[1]{{#1}}
\newcommand{\df}[1]{{#1}}

\url{www.pnas.org/cgi/doi/10.1073/pnas.0709640104}
\copyrightyear{2013}
\issuedate{Issue Date}
\volume{Volume}
\issuenumber{Issue Number}

\begin{document}

\title{Optimizing water permeability through the hourglass shape of aquaporins}
 
\author{Simon Gravelle\affil{1}{Institut Lumi\`ere Mati\`ere, UMR5306 Universit\'e Lyon 1-CNRS, Universit\'e de Lyon 69622 Villeurbanne, France}, 
Laurent Joly\affil{1}{}\thanks{To whom correspondence should be addressed. Email: <laurent.joly@univ-lyon1.fr>}, Fran\c cois Detcheverry\affil{1}{}, Christophe Ybert\affil{1}{}, C\'ecile Cottin-Bizonne\affil{1}{} \and Lyd\'eric Bocquet\affil{1}{}}

\contributor{Submitted to Proceedings of the National Academy of Sciences
of the United States of America}

\maketitle

\begin{article}
\begin{abstract}
The ubiquitous aquaporin channels are able to conduct water across
cell membranes, combining the seemingly antagonist functions of a very
high selectivity with a remarkable permeability. %in view of their subnanometric size. 
While molecular details are obvious keys to perform these tasks, the overall efficiency of transport in such nanopores is also strongly limited by viscous dissipation arising at the connection between the nano-constriction and the nearby bulk reservoirs.
In this contribution, we focus on these so-called entrance effects and
specifically examine whether the characteristic hourglass shape of
aquaporins may arise from a geometrical optimum for such hydrodynamic
dissipation. Using a combination of finite element calculations and
analytical modeling, we show that conical entrances with suitable
opening angle can indeed provide a large \red{increase of the overall channel
permeability.} Moreover, the optimal opening angles that maximize the
permeability %with respect to entrance effects 
are found to compare well with the angles measured in a large variety
of aquaporins.
This suggests that the hourglass shape of aquaporins could be the
result of a natural selection process toward optimal hydrodynamic
transport. 
% On the whole, also
\blue{Finally, in a biomimetic perspective, these results provide guidelines to design artificial
nanopores with optimal performances.} 
 % \textit{It is therefore striking that among various constraints, aquaporins shape has evolved toward an hourglass geometry that optimizes their hydrodynamic permeability.} 
 % \textit{PHRASE PRECEDENTE A VOIR: LAISSER, SUPPRIMER, REMPLACER PAR CELLE DE LAURENT?"It is therefore tempting to interpret the hourglass shape of aquaporins as the result of an evolution-driven optimization of their hydrodynamic permeability".}
\end{abstract}

\textit{Aquaporin channels are able to selectively conduct water across cell membranes, with a remarkable efficiency. While molecular details are crucial to the pore performance, permeability is also strongly limited by viscous dissipation at the entrances. Could the hourglass shape of aquaporins optimize such entrance effects? We show that conical entrances with suitable opening angle can indeed provide a large increase of the channel permeability. Strikingly, the optimal opening angles compare well with the angles measured in a large variety of aquaporins, suggesting that their hourglass shape could be the result of a natural selection process toward optimal permeability. This work also provides guidelines to optimize the performances of artificial nanopores, with applications in desalination, ultrafiltration or energy conversion.\\} 

\keywords{nanofluidics | hydrodynamic permeability | biochannels | aquaporin}

%\abbreviations{AQP, aquaporin; BC, boundary condition; FE, finite element}

\dropcap{A}quaporins (AQP) are water selective channels, ubiquitous in the living world 
\cite{Borgnia1999,Agre2004}. They are
involved in many physiological processes and play  
a crucial role in water exchanges across membranes, in particular under osmotic gradients. 
%This class of proteins is widespread in the living world and 
AQP are considered to be excellent water filters, able to achieve
contradictory and exquisite tasks: they exhibit high water
permeability, while ensuring excellent water selectivity
\cite{Murata2000,Sui2001}. 
From a technological point of view, designing artificial systems with similar performances would make a breakthrough
with applications in water desalination, ultrafiltration or energy harvesting based on osmotic power.  
%back to AQP While water selectivity is provided by a subtle and specific structure in a narrow constriction, 
%Aqueous pathway
By overcoming the limits of macroscopic transport, novel nanofluidic systems -- involving transport of fluids in specifically designed nanochannels -- have recently raised great hopes towards the realization of such achievements \cite{Rasaiah2008,Bocquet2010}.
% , {\it i.e.} the exploration of fluid transport at the nanoscales,
%has allowed to idendity new
One may quote in particular experiments reporting
%Flows at the nanoscale can exhibit features radically different from their macroscopic counterpart. 
%A case in point is the nearly frictionless transport of water inside carbon nanotubes.
%First suggested by experiments reporting unexpectedly 
high flow rate of water through carbon nanotube membranes \cite{Majumder2005,Holt2006}, 
%this surprising behavior, and its molecular mechanism, are now theoretically understood~\cite{Falk2010}. 
as well as giant osmotic effects in boron nitride nanotubes \cite{Siria2013}.

Yet, in order to design artificial nanopores with optimized performances, it would be interesting to investigate more closely the 
solution reached by biological nanochannels to perform similar functionalities. An intriguing aspect of AQP, highlighted in Fig.\,\ref{fig:AQP}, is the hourglass
shape exhibited by the aqueous pathway, with the inner part of the pore connected to the bulk water reservoirs via cone-shaped
vestibules. The central pore is of molecular size, and water is
transported in this region as a single file \cite{Hummer2008}. Selectivity is achieved
in this molecular scale confinement via a subtle molecular
organization of the confining pore
\cite{Murata2000,Sui2001,Groot2001,Jensen2002,Schulten2004,Ho2009}. 
%\red{UPDATE REFS?}.
%\blue{\st{Furthermore, water dynamics in the single-file process has been
%associated with very low dissipation} \cite{Hummer2001,Falk2010}.} 
On
the other hand, the conical entrances are of much larger size, making
a slow transition towards the bulk \lb{water}. While the inner part of the
pore is ruled by \lb{one-dimensional molecular transport} \cite{Finkelstein}, continuum hydrodynamics should
apply to some extent in these cone-shaped vestibules, as it was shown
that the Navier-Stokes equation remains valid down to extremely small
length scales \blue{(typically 1\,nm for water)} \cite{Bocquet2010,Thomas2009}. In this simplifying picture, the
central part is the key to ensure the selectivity %\lb{\st{and high
                                %permeability}} 
at the molecular scale, but the overall transport efficiency nonetheless incorporates a priori effects occurring in the conical entrances, that makes the transition to the bulk.

Accordingly, we raise in this paper the question of hydrodynamic entrance effects.
Specifically, we study whether the hourglass shape of AQP corresponds to an optimum with respect to hydrodynamic dissipation.
As we demonstrate below, the hourglass geometry does indeed display an optimal angle for which entrance effects are minimized, and the resulting gain in overall channel permeability can reach hundreds of percents.
Strikingly, geometrical parameters measured on a variety of AQP are found to compare well with these optimal opening angles.
In a broader biomimetic perspective, our findings
point to general design rules to minimize entrance effects and ensure
optimal transport in artificial nanopores.
% Accordingly, we raise in this paper the question of hydrodynamic
% entrance effects. More precisely, we study whether the hourglass shape
% of AQP corresponds to an optimum with respect to hydrodynamic
% dissipation, therefore best responding to the necessity towards a
% global fast water transport. \blue{In a broader biomimetic perspective, this
% raises the question of the general rules to design artificial
% nanopores in order to minimize dissipation in the entrance regions and
% maintain optimal transport and permeability.} As we demonstrate below,
% the hourglass geometry does indeed display an optimal angle for which
% entrance effects are minimized, and the resulting gain in overall
% channel permeability can reach hundreds of percents. Strikingly,
% geometrical parameters measured on a large variety of AQP are found to
% compare well with these optimal opening angles.

\section{Hydrodynamic Entrance effects}

\begin{figure*}
\centerline{\includegraphics[width=0.8\linewidth]{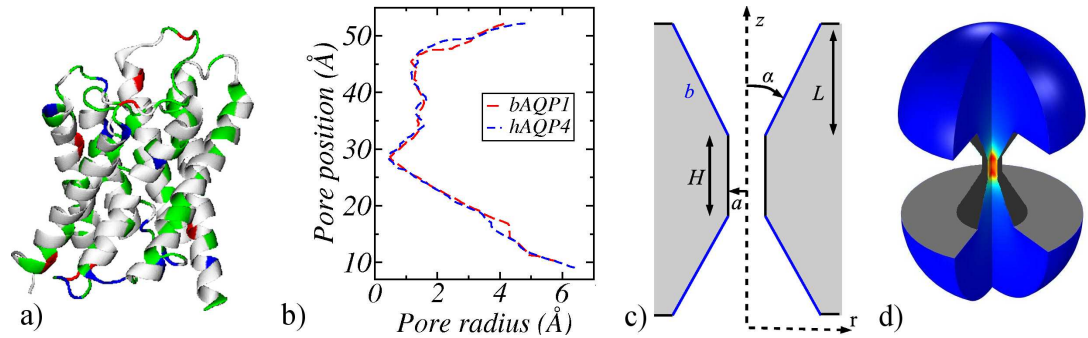}}
\caption{a) Molecular structure of human aquaporin 4 (hAQP4) obtained
  from the Protein Data Bank \cite{Bernstein1977}.
%and plotted with VMD \cite{Humphrey1996}.  
b) Profiles of two aquaporins collected from
Ref. \cite{Ho2009}. The pore dimensions were estimated using the HOLE program \cite{Smart1996}. 
%c) Axisymmetric representation of the cavity running through the hAQP4. 
c) The biconical channel considered in this work. The  central
cylinder, where perfect slip is assumed, is connected to two truncated
cones of length $L$, opening angle $\alpha$ and with perfect or finite
slip (slip length $b$, see text). 
d) Schematic of the system used for the finite element (FE) resolution of the Stokes equation (COMSOL). 
The channel is connected to half-spherical reservoirs (not to scale). 
Color represents the magnitude of fluid velocity, from blue (slow flow) to red (fast flow).}
\label{fig:AQP}
\end{figure*}

The question of entrance effects in hydrodynamic flows goes back to
the early history of hydrodynamics.
In 1891, Sampson obtained the exact solution for the Stokes flow through a circular aperture in an infinitely thin membrane \cite{Sampson1891}
%As a by product, he 
and found that the pressure drop \blue{$\Delta p$} across the membrane was
\begin{equation}
\Delta p = \frac{3 \eta}{a^3} \times Q , 
\label{eq:sampson}
\end{equation}
with $a$ the aperture radius, $\eta$ the liquid dynamic
viscosity and $Q$ the flow rate through the aperture.
The pressure drop results from the narrowing of the streamlines over a
range $a$. The scaling results accordingly from Stokes' equation:
$\eta \Delta v = \nabla p \Rightarrow \eta v/a^2 \sim \Delta p/a$,
with $v \sim Q/a^2$ the typical fluid velocity.
Extending this estimate to the converging flow into a cylindrical pore, the previous Sampson's formula
provides a very good estimate of the access pressure drop \cite{Weissberg1962}, as highlighted by an exact calculation \cite{Dagan1982}.
% to a very good approximation the a similar entrance contribution
%shows up
%A similar contribution shows up for the flow entering The same 
%Later, Dagan et al investigated in detail the Stokes flow through 
%a cylindrical channel of finite length~\cite{Dagan1982}. 
%With the exact solution at hand, 
%they showed that the total resistance is very well approximated by summing two contributions: 
%that of a Poiseuille flow assumed to hold through the entire tube length, even if this does not hold in the vicinity of the channel end, 
%and an additional access resistance, given by Sampson's formula. 

%When viscous dissipation inside the pore is extremely small (e.g. in carbon
%nanotubes or AQP), such end effects become the limiting factor for hydrodynamic
%transport, as shown recently \cite{Sisan2011,Nicholls2011}, 
%and exemplified by the capillary uptake in carbon nanotubes
%\cite{Joly2011}.
\lb{Now, if one consider water transport through a pore, the
  dissipation occurring in the bulk access regions yields an upper  
bound to the hydrodynamic permeability. %(thus an upper bound to the hydodynamic permeability).
}
\lb{
%End effects are all the more important that the viscous dissipation inside the pore itself decreases. 
%This is especially the case i
In nanochannels  where inner dissipation is extremely small 
(such as carbon nanotubes \cite{Majumder2005,Holt2006,Hummer2001,Falk2010}), 
entrance effects thus act as the limiting factor for water
transport, as shown recently
\cite{Sisan2011,Nicholls2011,Walther2013},  and exemplified by the
capillary uptake of water in carbon nanotubes \cite{Joly2011}.} 
\df{ For AQP, whose dissipation in the inner part has not been fully characterized,
a simple estimate nevertheless shows} 
\lb{that entrance effects contribute to a large part to the global hydrodynamic resistance. Indeed, }
\blue{the permeabilities of AQP reported in the literature vary
typically in
the range $p_f = 0.5$ to $1.5\times 10^{-19}$\,m$^3$/s \cite{Hashido2007}.
%
%\red{Citations ?}
\lb{This is to be compared to a typical order of magnitude for the entrance
contribution to the permeability\footnote{
The permeability $p_f$ is defined in the physiology literature from the
flux of water $\Phi_w$ (in moles per unit time) resulting from an osmotic pressure difference $\Delta \Pi$ across the pore, as
$p_f=\Phi_w \mathcal{R} T/\Delta \Pi$, with \df{$\mathcal{R}$} the gas constant \cite{Finkelstein}. It is related to the hydrodynamic permeability $K$ defined as the ratio of the flow rate and pressure drop, as $p_f=K \times \mathcal{R} T/\mathcal{V} _w$, with \df{$\mathcal{V} _w$} the molar volume of water.}}, 
which can be estimated from Sampson's formula\footnote{
Applying Sampson formula at the subnanometric level may seem a bold approximation. 
However, in Ref.~\cite{Suk2010}, Suk and Aluru have examined the single-file flow of water 
through a graphene sheet pierced with a subnanometric hole. % $0.78$ nm in diameter. 
Their molecular dynamics simulation leads to a hydrodynamic permeability $K=3\,10^{-27}$\,m$^3$/(Pa\,s). 
To compare to Sampson's formula, we need the fluid viscosity and the orifice radius $a$. 
The former is, for the SPC/E model at 298\,K,
$\eta=0.73\,10^{-3}$\,Pa\,s \cite{AngelGonzalez2010}. 
To estimate the latter consistently with Fig.\,\ref{fig:AQP}a, 
we follow the convention of the HOLE program, i.e. $a=d/2-r_{C,vdW}$, 
where $d=0.75$\,nm is the diameter obtained using the center-to-center distance of carbon atoms, 
and $r_{C,vdW}=0.17$\,nm is the carbon Van der Waals radius.  
Equation\,\eqref{eq:sampson} then yields a hydrodynamic permeability $K=3.9\,10^{-27}$\,m$^3$/(Pa\,s). 
This suggests that Sampson's formula gives the correct order of magnitude even at the subnanometric level, 
down to the single-file regime.}, Eq. \eqref{eq:sampson}, giving
$p_f ^{{\rm access}}= 1.5\times 10^{-19}$\,m$^3$/s. 
\blue{This estimate shows that entrance effects contribute importantly to the overall AQP permeability and cannot be overlooked}  
.}
%, and as such it is important to consider them.} 
\lb{It is therefore interesting to investigate whether -- regardless of the dissipation occurring in the central, molecular part of the pore --, strategies exist to reduce significantly these entrance effects.} 
\lb{Altogether, the global permeability of a pore may be written as two resistances in series, 
\begin{equation}
p_{f,tot}^{-1}=p_{f,{\rm access}}^{-1}+p_{f,central}^{-1}
\label{permeability}
\end{equation}
where the two contributions in the right hand side stand for the two
access resistances and the resistance originating in the central part
of the channel.} 
%
% \st{In practice, streamlines must be curved to enter the tubes, and velocity gradients will always exist in the entrance region. However connecting directly a
% cylindrical tube to the reservoir induces an abrupt transition and the question remains whether a better strategy to bend the streamlines could be found.}
%
\lb{In the following we focus on the first, access part of the permeability, so that our result constitute an upper bound to the total permeability. We show that
a conical shape allows to reduce considerably the corresponding entrance dissipation, and thus provides an optimal shape for water transport.}

\section{Optimized permeability of hourglass-shaped pores}

Inspired by the AQP shape (Fig.\,\ref{fig:AQP}), we investigate
hydrodynamic entrance effects in nanopores with an hourglass shape.
We consider a pore made of a central cylinder
%here we assume a frictionless in the center, to mimick
%Figure~\ref{fig:AQP} illustrates the four steps taken in building a minimal model of entrance effects within the continuum hydrodynamic description.
%The starting points are the molecular structures obtained from high-precision X-ray cristallography, 
%which are available for several aquaporins. 
%From this wealth of information, we retain only the radius profile of the channel, 
%as estimated by the Hole program.
%Though irregular, those profiles can be divided into three parts:
%a central constriction,
%and two diverging portions.  
%As detailed in the methods section,
with radius $a$ and length $H$, connected to two conical vestibules, 
with length $L$ and opening angle $\alpha$ (Fig.\,\ref{fig:AQP}c). 
\red{This system will be studied at the level of the continuum Stokes equation.
Given that aquaporins dimensions are subnanometric, with the pore
mouth in the nanometer range down to a central pore diameter $\sim 0.3$\,nm,
using continuum hydrodynamics may certainly be questioned. 
However, it is known that Navier-Stokes equation is remarkably robust,
remaining valid down to the nanometer scale
\cite{Bocquet2010,Thomas2009}, so that it should apply -- at least to
some extent -- in the entrance regions connecting the \lb{pore mouth to the bulk}}\footnote{\df{See also footnote~2 above.}}. 
% \blue{\st{Furthermore, if not quantitative in their predictions, 
% hydrodynamic continuum equations remain the most convenient and useful  framework to investigate generic features of the fluid flow, 
% independently of any specific molecular structure.}} 

Now, fluid transport inside the central part of the aquaporin belongs
to the single-file regime and
the physics at play here cannot be captured by a continuum description. 
However, we are not interested here in the specific selectivity of the
AQP -- which would indeed require a detailed atomic modeling \cite{Groot2001,Jensen2002,Schulten2004,Ho2009}. 
%and a simplifying description can still be proposed.
%to account for the low friction expected in this section, we assume perfect-slip boundary condition. 
\lb{In order to focus on entrance effects \df{alone}, we consider a simplified view in which all
  dissipation in the AQP central channel is neglected. This is done by assuming a {\it
  perfect-slip} boundary condition on the pore surface (see Methods), so that surface friction is vanishing in this central part. By doing so,
we thus focus on the entrance contribution to the permeability, {\it i.e.} the first term in Eq.(\ref{permeability}).
Therefore, our results provide an upper bound to the total permeability of the pore.}
% \red{In order to mimic the transport in this central part, and in
% particular the expected negligible friction, we assume a {\it
%   perfect-slip} boundary condition (see Methods) in the continuum
% description, so that surface friction is vanishing in the central part of the pore.} 

Boundary conditions (BC) in the conical regions have \lb{also} to be prescribed (see Methods). As shown to be relevant for nano-scale flows \cite{Bocquet2007}, we assume a partial slip BC on the cones' walls, with the  velocity field at the surface obeying $b\,\partial_n v_t\vert_{\rm surf} = v_t\vert_{\rm surf}$,
%Besides the channel geometry, 
%we need to specify the boundary conditions. 
%%
where $b$ is the so-called slip length and $v_t$ is the tangential
component of the velocity. Interestingly, the molecular structure of
the AQP in contact with water is mostly hydrophobic
\cite{Murata2000,Sui2001,Rasaiah2008} \blue{(with hydrophilic patches to ensure that
water penetrate through the pore)}, and in line with recent work on
hydrodynamic slippage \cite{Bocquet2007,Bocquet2010}, a slip length in the range
of tens of nanometers may typically be expected. This value is large
as compared to the other typical length scale of the nanopore, $b\gg
a$. Consequently, we will start our discussion by assuming perfect slip BC ($b=\infty$), and then relax this condition in a second step.
%For now, perfect slip is also assumed at the cone wall.
%Below, we will relax this assumption and consider a finite slip length.
%Second, as regards the infinite slip length, 
%we note that hydrophobic surfaces lead to slip length in the tens of nanometer, in effect much larger than aquaporins radii. 
The Stokes equation with the above BC is solved numerically, see
Methods. We will also present a simple model mimicking the generic
aspects of flow \lb{in the access regions of} AQP, which is able to provide the main features of hydrodynamic entrance effects. %, though not specific

\begin{figure}[h]
\centerline{\includegraphics[width=0.7\linewidth]{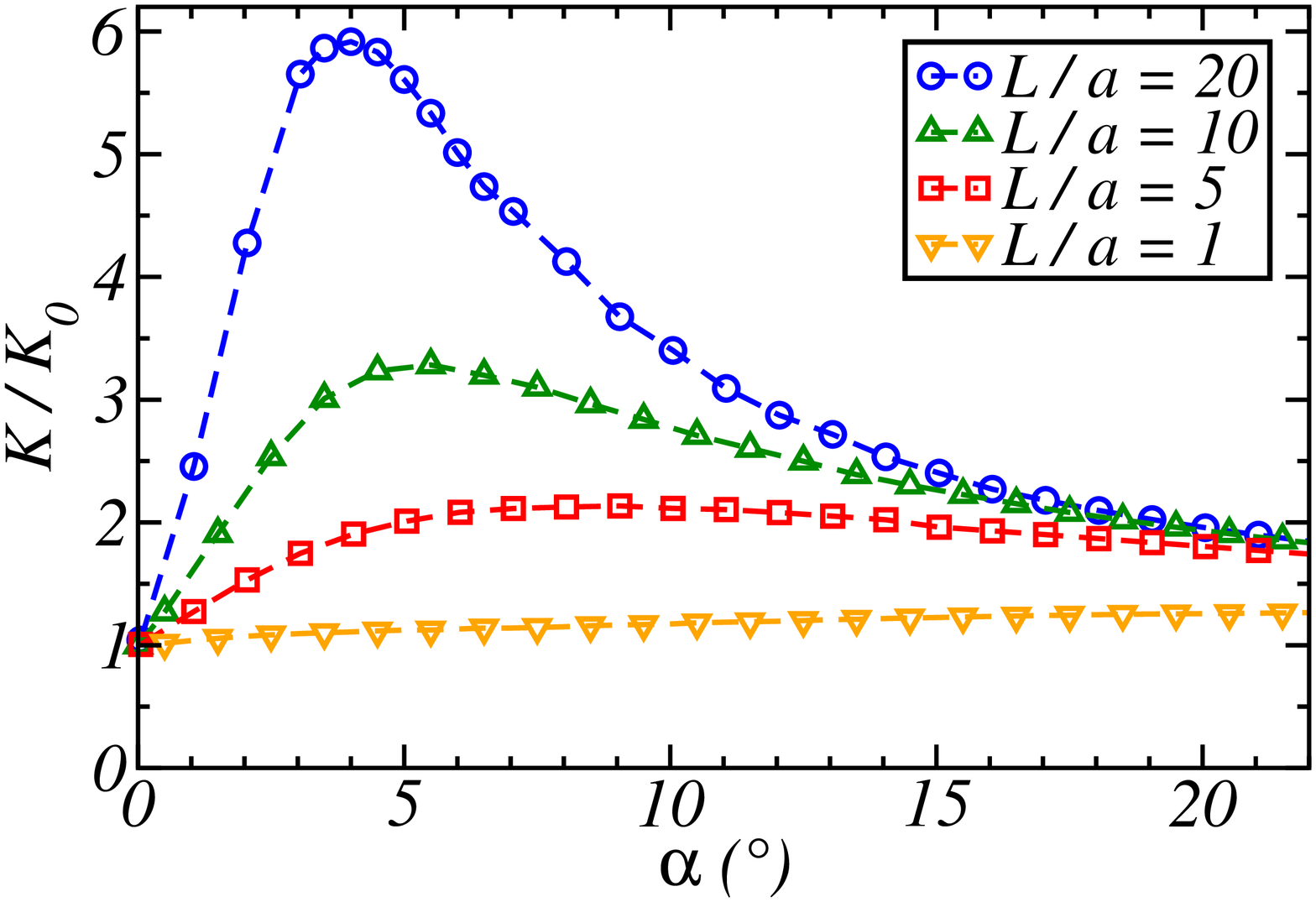}}
\caption{%a) 2D plan of the axisymmetric system used for finite element calculation. The central part of the nanochannel has got a radius $a$, a length $H>a$ and perfect slip boundary condition along it's wall. Two truncated cones with angle $\alpha$ and length $L$ are connected to the central part. Various slip boundary condition characterized by a slip length $b$ can be applied on cone's walls (blue lines on the drawing). 
Hydrodynamic permeability $K=Q/\Delta p$ of the hourglass
nanochannel as a function of the opening angle $\alpha$, obtained from
FE calculations. $K$ is normalized by $K_0=K(\alpha=0)$.
Perfect slip ($b=\infty$) is assumed on the cones inner walls. 
Each curve corresponds to a cone length $L$.}
\label{fig:draw_curve}
\end{figure}

\red{Our main result is illustrated in Fig.\,\ref{fig:draw_curve},
which shows the hydrodynamic permeability $K$ of the hourglass channel 
as a function of the opening angle $\alpha$. 
The permeability $K = Q/\Delta p$ provides the flow rate $Q$ for a
given pressure drop $\Delta p$.} 
%The resistance $R=\Delta p/Q$ gives the pressure drop $\Delta p$ obtained upon imposing a flow rate $Q$, 
%it was computed by solving Stokes equation numerically (see Methods). 
%
As highlighted in this figure, for any cone length $L$, the
permeability is a {\it non-monotonic} function of the opening angle of
the pore: starting from the cylinder geometry ($\alpha = 0$), the
permeability starts by increasing very quickly with $\alpha$, before decreasing slowly for larger angles.
%A slight variation of the angle $\alpha$ towards values around 5-10$^\circ$ leads to an abrupt increase of the permeability, before decreasing again slowly for larger angles.
%For all cone lengthes $L$ considered, 
%
There is accordingly an optimal angle $\alopt$ which maximizes the channel permeability, 
{\it i.e.} yields a maximal flow rate under a given pressure forcing.
Compared to the cylindrical case ($\alpha=0$), 
the optimal geometry ($\alpha=\alopt$) yields a very significant increase in permeability, especially for long cones. 
At $L/a=20$ for instance, the optimal permeability is 6 times larger
than the one of a cylinder. 
Although it increases for shorter cones,
the optimal angle remains small, below 10$^\circ$ for
$L/a>5$. Surprisingly, a tiny departure from the straight cylinder
makes for a large effect on \blue{entrance} dissipation. 
This is an unexpected result and in order to gain insight into its origins, 
%To rationalize those results, 
we now develop a simplified model to rationalize viscous dissipation in the hourglass channel.

\section{Towards a simplified analytical model}

\begin{figure}[h]
\centering
\includegraphics[width=0.8\linewidth]{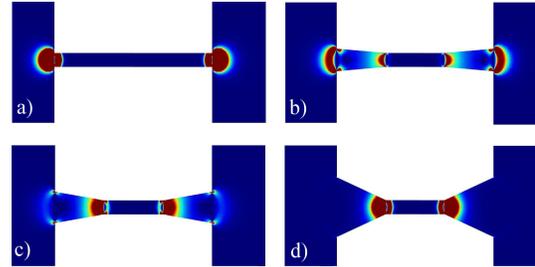}
\caption{Local viscous dissipation rate $\disrate$ (see text) inside the nanochannel for different values of the angle $\alpha$.  
The color scale, from blue to red, indicates increasing values of local viscous dissipation. Perfect slip boundary condition is imposed on the cone walls. From a) to d), $\alpha=0,5,10$ and $25^\circ$.}
\label{fig:cone_dissipation}
\end{figure}

So far,  we assumed in a first step a negligible friction on the cone's surface (perfect slip). In this situation, the dissipation is expected to occur mostly within the two transition regions: from the reservoir to the cone (first entrance), and 
from the cone to the cylinder (second entrance). 
This is confirmed by the numerical results, as highlighted in Fig.\,\ref{fig:cone_dissipation}, 
where we have plotted the local viscous dissipation rate $\disrate=2
\eta \boldsymbol{\Delta\!:\!\Delta}$, with 
$\boldsymbol{\Delta}=  \left[ \boldsymbol{\nabla \vcv} + \boldsymbol{\nabla \vcv} ^T   \right]/2$ the strain rate tensor. The spatial extent of both regions is given by the local radius, 
with a prefactor close to unity. This figure shows that increasing the 
opening angle shifts the dissipation from the first entrance to the second. This is to be expected if one realizes
that the streamlines have basically to follow the surface of the pore, and that the
second angle between the cone and the cylinder increases as the first
angle between the wall and the cone decreases
(the sum of the two angles being constant due to geometry).

This picture suggests to describe the total hydrodynamic resistance of
the pore, $R=K^{-1} = \Delta p / Q$ (the inverse permeability), as the sum of the
various contributions (channel entrance, cone region, and cylinder 
entrance\,\footnote{Note that the three resistances do not identify with dissipation inside the volume of reservoir, cone and cylinder respectively. For example, the entrance resistance includes dissipation taking place both in the reservoir and in the cone.}) in series, 
%It is customary in microfluidics to evaluate the total impedance of a system  from the impedance of each components, 
as for a resistive circuit:
%Such an approach is justified if the system can be separated into several independent parts, 
%each with a well-defined impedance. 
%That this is the case in the biconical channel can be seen in Fig.~\ref{fig:cone_dissipation}, 
%where we have plotted the local viscous rate $\disrate=2 \eta \Delta : \Delta$, where $\Delta=  \left[ \nabla \vcv + \nabla \vcv ^T   \right]/2$ is the strain rate  tensor.  
%Dissipation appear mainly localized in two regions:
%the entrance to the cone and the entrance to the cylinder. 
%The spatial extent of both regions is given by the local radius, 
%with a prefactor close to unity. 
%
%For a cone long enough, typically $L/a \geqslant 5$, 
%the two main dissipative regions are sufficiently far away to be considered independent. 
%Accordingly, we write the total channel resistance as a sum of three contributions:
\begin{equation}
R = \Rent + \Rcone + \Rentcyl. 
\label{eq:Rsum}
\end{equation}
For such a decomposition to hold, 
both the cylinder radius $a$ and the entrance radius $a'=a+ L \tan \alpha $ 
should remain small compared to $L$, which is valid 
%namely $a+a' < L$ at the very least.  
%This implies not only a 
for large $L/a$ ratio
and small opening angle $\alpha$. %remains small. 
%We will use repeatedly those two assumptions in the following, 
%as we try to evaluate each contribution to the resistance.  

In a cone of infinite extent with arbitrary opening angle and
perfect slip at the wall, 
the Stokes flow is purely radial with a velocity that decreases as $1/r^2$, where $r$ is the distance from the apex. 
One may then verify that the pressure drop, evaluated from $\nabla p =
\eta \Delta \vcv$, vanishes in this situation.
Accordingly, for the case $b=\infty$ that we consider so far, 
$\Rcone$ is thus negligible, in agreement with numerical results, see Fig.\,\ref{fig:cone_dissipation}. 

Now, to proceed further and estimate the remaining contributions in Eq.\,\eqref{eq:Rsum}, we need to estimate
entrance hydrodynamic resistance for two configurations: (i) a conical
aperture with a finite angle and perfect slip; (ii) a cone-to-cylinder entrance.
These are generalized Sampson geometries, which we consider now.

\begin{figure}[h]
\centerline{\includegraphics[width=0.9\linewidth]{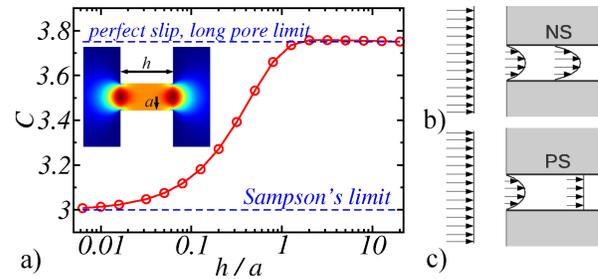}}
\caption{ a) $C = R a^3/\eta$ (see text) as a function of the pore
  length $h$ for a perfectly slipping cylindrical channel, obtained with FE calculations (points).   
The red line is a guide for the eyes. 
Inset: Velocity field in the considered channel. 
Color represents the amplitude of the fluid velocity, from blue (slow flow) to red (fast flow). 
b,c) Schematic profiles of axial velocity in a channel with no slip (NS) and perfect slip (PS) boundary condition. 
%{\color{red} Color scale from blue to red symbolizes the dissipation from low to
%high (enlever le code couleur, perturbant? LJ: JE SUIS D'ACCORD)}. 
From left to right: outside far field, entry profile and inside far field.}
\label{fig:entrance}
\end{figure}

\subsection{Generalized Sampson formula (i): aperture with perfect slip}

To evaluate $\Rent$, we build on previous works. 
%In 1891, Sampson obtained the exact solution for the Stokes flow through a circular orifice in an infinitely thin wall ~\cite{Sampson1891}. 
%As a by product, he found that the hydrodynamic resistance was
%\begin{equation}
%\frac{\Delta p} {Q }= \frac{3 \eta}{a^3},
%\label{eq:sampson}
%\end{equation}
%where $a$ is the orifice radius and $\eta$ is the liquid dynamic viscosity. 
%Later, Dagan et al investigated in detail the Stokes flow through a cylindrical channel of finite length~\cite{Dagan1982}. 
%With the exact solution at hand, 
%they showed that the total resistance is very well approximated by summing two contributions: 
%that of a Poiseuille flow assumed to hold through the entire tube length, even if this does not hold in the vicinity of the channel end, 
%and an additional access resistance, given by Sampson's formula. 
%
\comm{We neglect the effect of the conical shape
of the aperture and evaluate $\Rent$ for the geometry of a flow
entering into a cylinder, an approximation valid for long cones with small angle.}
%In writing this expression, we assume that the cone is sufficiently long and that the cone angle is small
%(so that angle dependency of $C$ is omitted).   
%it is clear that  $\Rent$ should increase with $\alpha$. 
%\comm{As a first approximation, 
% Since most of the viscous dissipation occurs
% outside the channel due to the bending of the streamlines, from the reservoir to the aperture, this
% approximation should not bring significant differences.}
%
As quoted above, Sampson calculation was generalized by Dagan
\textit{et al.} to such geometry
%e geometry of a flow entering into a cylinder
\cite{Dagan1982}. However these calculations assumed a no-slip BC at
the pore walls and we need first to generalize
this result to perfect slip as considered here. 
On dimensional grounds, the access resistance of the finite tube is expected to write 
$R=\Delta p / Q = C \eta/a^3$, where $C$ is a numerical prefactor
($C=3$ in Sampson expression, Eq. \eqref{eq:sampson}).
We have computed numerically this prefactor $C$ using FE calculations
for a perfectly slipping cylinder of finite length, Fig. \ref{fig:entrance}a.
%Since those results apply only to the channel with no-slip boundary condition, 
%they need to be extended to the perfect slip case. 
%Rather than  being Poiseuille-like, the flow in the interior of a long tube is now plug-like, 
%without dissipation, making the tube length irrelevant. 
%Accordingly, on dimensional grounds, the resistance of the finite tube reads as in Eq.~\ref{eq:sampson}, 
%$ R=C \eta/a^3$, where $C$ is a numerical factor. 
%Figure~\ref{eq:sampson} shows this $C$ coefficient, obtained from FE calculations, as a function of tube length $h$. 
%
For very short tubes, 
the BC at the inner wall becomes irrelevant, and Sampson's result for
the infinitely thin membrane is recovered with $C=3$.  
As the tube gets longer, however, 
the $C$ coefficient increases up to a plateau value of
$C_\infty \approx 3.75$, 
implying that switching from no-slip to perfect slip inside the tube yields a 25\% {\it increase} in the access resistance. 
Though counter-intuitive at first sight, 
this behavior can be qualitatively understood by examining the flow profile at the channel end. 
In the no-slip case (Fig.\,\ref{fig:entrance}b),
 this ``entry'' flow profile is, to a very good approximation, halfway between the parabolic profile of Poiseuille flow 
and the elliptic profile found in Sampson's solution \cite{Dagan1982}.  
As a result, the transition to a plain parabolic profile inside the channel involves only a small dissipation. 
Because the velocity must vanish at the corner, 
the entry profile in the perfect-slip tube is quite similar to the no-slip case, 
suggesting a comparable amount  of viscous losses outside the tube. 
Now,  inside a tube with perfect slip at the boundaries, 
the transition to a plug profile requires significant reorganization
of the streamlines (Fig.\,\ref{fig:entrance}c), 
resulting in a higher dissipation, hence a higher $C$. 
As a side remark, we find that the BC on the external wall has a negligible impact on the access resistance, presumably because any slip that could happen there is strongly hampered by the vanishing velocity at the corner. 

\df{Coming} back to the case of the hourglass-shaped pore, we now estimate the
hydrodynamic resistance of the first entrance  as
\begin{equation}
\Rent= \frac{C_\infty\, \eta}{a'^3}, \quad a'=a+ L \tan \alpha, 
\label{eq:Rent}
\end{equation}
with $C_\infty=3.75$.

\subsection{Generalized Sampson formula (ii): cone-to-cylinder entrance}

%It remains to evaluate $\Rentcyl$. 
No prior work seems to have investigated the Stokes flow at the
junction between a cone and a cylinder, 
both with perfect slip \footnote{This is not surprising since until recently, the perfect-slip BC was not deemed physically relevant.}. We could not obtain an analytical solution either for this geometry.
However numerical calculations (see Methods) suggest that the hydrodynamic resistance at a cone-to-cylinder transition can be well approximated by 
\begin{equation}
\Rentcyl= C_\infty\, \sin \alpha\, \frac{\eta}{a^3} , 
\label{eq:Rentcyl}
\end{equation}
with again $C_\infty=3.75$.
%An exact solution, if possible at all, would anyway involves numerical solving (see appendix). 
This behavior can be rationalized on the basis of the following
%We resort instead to a 
``back-of-the-envelope'' argument. %to find the functional dependence of   $\Rentcyl$ on the opening angle $\alpha$. 
Far from the junction, streamlines are parallel in the cylinder and radially divergent in the cone. 
Dissipation occurs only in the vicinity of the junction, where the
streamlines change direction by an angle $\alpha$, so that $\nabla v \sim v_0 \sin \alpha/a$.
The pressure drop $\Delta p$ is then given roughly as $\Delta p \sim \eta \nabla v  \approx \eta v_0 \sin \alpha/a $, pointing to the origin of the $\sin \alpha$ dependence  for  $\Rentcyl$.
Now in the case $\alpha=\pi/2$, one should recover the resistance
$\Rent$ with prefactor $C_\infty$, leading to Eq. \eqref{eq:Rentcyl}.
%this 
%leading to a similar $\sin \alpha$ dependence  for  $\Rentcyl$ 
%\footnote{I dont follow this argument: why not say  $\nabla v \sim \sin \alpha$ and $\disrate \sim  \sin^2 \alpha$ ?}. 
%Dimensional analysis again leads to  $\Rentcyl=C' \eta \sin \alpha /a^3$. 
%The numerical coefficient $C'$ is obtained by taking the particular case $\alpha=\pi/2$, for which one recovers the resistance $\Rent$ computed above~\footnote{As noted above, the boundary condition on the external wall has negligible influence on the C coefficient.}. 
%We than expect
% \begin{equation}
%\Rentcyl= \frac{3.75 \eta}{a^3} \sin \alpha.  
%\label{eq:Rentcyl}
%\end{equation}
%FE calculations indicate that Eq.~\ref{eq:Rent} provides a good approximation to the exact result, 
%with a relative differnce that is at most XX\%. 

\subsection{Total hydrodynamic resistance}
\begin{figure}[h]
\centering
\includegraphics[width=0.7\linewidth]{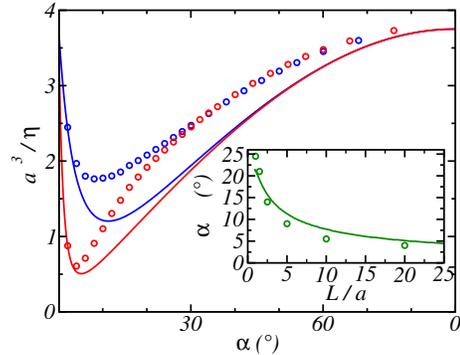}
\caption{Pore resistance $R$ versus cone angle $\alpha$ for perfect
  slip in the cones: comparison between FE calculations (circles) and
  Eq. \eqref{eq:Rmodel} (lines). Results are presented for two pore
  lengths: $L/a=20$ (red) and $L/a=5$ (blue).  
Inset: optimal angle $\alpha_{opt}$, for which the resistance is
minimized, as a function of cone length.}
\label{fig:comparison1}
\end{figure}

Collecting Eqs. \eqref{eq:Rsum}, \eqref{eq:Rent} and
\eqref{eq:Rentcyl}, and remembering that $\Rcone \approx 0$, yields the total resistance of the hourglass channel in our simplified model:
\begin{equation}
R= \frac{C_\infty\eta }{a^3} \left[    \left(1+\frac{L}{a} \tan \alpha \right)^{-3} + \sin \alpha \right].
\label{eq:Rmodel}
\end{equation}
This relation exhibits a non-monotonous behavior versus the angle
$\alpha$, as shown in Fig. \ref{fig:comparison1}:
%It is now clear why a minimum appears, since 
the first term in the right-hand-side decreases rapidly with $\alpha$, 
while the second term steadily increases. 
\comm{Physically, these two terms account for dissipation at the first and
second entrance respectively, and their variations confirm the
qualitative picture illustrated in Fig.\,\ref{fig:cone_dissipation}.}
A minimum for the resistance -- thus a maximum for the permeability $K=R^{-1}$ -- is then found.
In particular, for long cones and small angles, $R \approx (\alpha
L/a )^{-3} + \alpha$, and the optimal angle decreases with the cone length  as $\alopt \sim (L/a)^{-3/4}$.  
More quantitatively, Fig.\,\ref{fig:comparison1} compares FE calculations and the predictions of our simplified model. While a quantitative agreement is not expected in view of the simplifying assumptions underlying our model, the latter is found to capture the optimization phenomenon. In particular, the variation of the optimal angle 
$\alopt$ versus length $L/a$ is well reproduced, see inset of Fig.\,\ref{fig:comparison1}.
%For the long conical section ($L/a=20$), the curves minima are close to each other, both in value and in optimal angle. 
%For short cones  ($L/a=5$), the assumption $L/a \gg 1$ breaks down and our simplified mode underestimates the resistance. 
%However, the prediction of the optimal angle remains rather accurate in the whole range $L/a\in[2,20]$, 
%even for the shortest cones considered (see inset of Fig.~\ref{fig:comparison}).  
%Note that even for $L/a=2$, $\alopt$ never exceeds 15$^\circ$, indicating that our assumption of small $\alpha$ is indeed justified. 

To summarize, there is an optimal angle which maximizes the permeability of an hourglass channel, and minimizes
viscous losses. This optimal angle takes rather small values, in the
range $\alopt \sim 5-20^\circ$ depending on the length of the conical vestibules,
%nanopore
but can lead to very strong improvements in the permeability. 
%showing that a slight adaptation 
%there is an optimal angle that minimizes viscous losses, and which gets smaller as the cone gets larger. 
\comm{Following Dagan \textit{et al.}, it should be possible to calculate the
exact solution for a flow inside a biconical nanochannel \cite{Dagan1982}.
We leave this ambitious work for future dedicated investigations.
However, an approximate model is presented in the Supporting Information.}
% This theoretical work, however interesting and ambitious as it
% is, should bring no more information than our simple model.
%in the spirit of the calculations carried on by Dagan \textit{et al.}

\begin{figure}[h]
\centering
\includegraphics[width=0.7\linewidth]{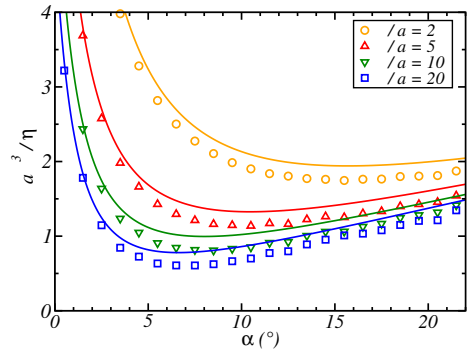}
\caption{
%Total resistance $\Delta p / Q$ versus opening angle $\alpha$ for fixed cone's length $L/a=5$ and for different values of slip length $b$ on cone's wall.
Pore resistance $R$ versus opening angle $\alpha$, for various slip
lengths $b$ in the conical regions: comparison between FE calculations
(symbols) and analytical expression (lines), see text for detail. The cone length is fixed to $L/a=20$.}
\label{fig:glissement_partiel}
\end{figure}

\subsection{Extension to finite friction on the cone surface}

\begin{figure}[h]
\centering
\includegraphics[width=0.9 \linewidth]{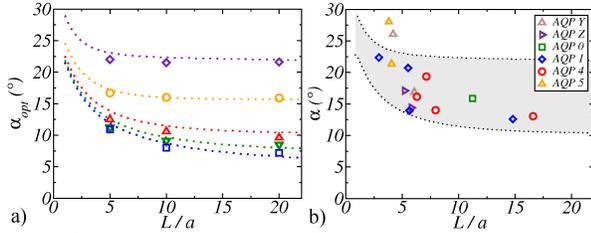}
\caption{a) Optimal angle as a function of cone length $L$ for
  various slip lengths (from top to bottom, $b/a = 1,2,5,10,20$): FE results (circles) and model (lines). 
b) Angle $\alpha$ evaluated in six aquaporins (see Methods). The gray
shaded area corresponds to model predictions for $b/a=1$ to $5$ (as shown in left panel).
Data are extracted from Refs. \cite{Ho2009,Cui2011,Groot2003,Harries2004,Zhang2013,Jensen2006,Fischer2009}.}
\label{fig:comparison}
\end{figure}

Up to now, we assumed a perfect slip BC at the cones' surfaces, 
  corresponding to the limit of large slip lengths compared to
  transverse dimensions $b/a\rightarrow\infty$. We now relax this
  assumption and consider finite $b/a$. 
Figure~\ref{fig:glissement_partiel} reports the results of numerical calculations for the hydrodynamic resistance
versus opening angle, for various slip lengths $b$ at the cone surface
and a fixed cone length, $L=20\ a$ (see Methods).
As shown in this figure, an optimal angle minimizing the hydrodynamic
resistance is still found for finite slip, and its value
increases with decreasing slip length, see also Fig. \ref{fig:comparison}a.

%Having treated the case of perfect slip at the wall, 
%we now examine the influence of solid-liquid friction, which is
%quantified by a finite slip length $b$ (see Methods).

Again a simplified model can be built. While we do not expect entrance effects to be radically modified, a supplementary dissipation will now occur due to finite slippage at the cone surface.
This contribution, $\Rcone$ in Eq. \eqref{eq:Rsum}, can be calculated within lubrication theory, valid for small angles $\alpha$, as
\begin{equation}
{\Rcone}= 2 \int_0^L dz\,\frac{8\eta }{\pi a(z)^4} \left( 1+\frac{4b}{a(z)}\right)^{-1}, %\qquad a(z)=a + z \tan \alpha. 
\label{eq:Rcone}
\end{equation}
with $a(z)=a + z \tan \alpha$ the local radius of the cone. Accordingly, an analytical expression for $\Rcone$ can be obtained, but its cumbersome expression is not particularly illuminating and we do not report it here.
%The total resistance  $\Rcone = \int_0^L \frac{d\Rcone}{dz} dz$ can be obtained analytically, but the complicated expression is not particularly illuminating. 
%
Gathering all contributions in Eq. \eqref{eq:Rsum} leads to an
analytical expression for the hydrodynamic resistance of the hourglass pore with finite slip length $b$ on the cones. 
%two changes happen.
%First, variation in $\alpha$ induce only weak variation in the resistance $R$.  
%Second, the optimal angle shifts to higher values. 
This expression is compared to the numerical calculations in Figs. \ref{fig:glissement_partiel} and \ref{fig:comparison}a.
Altogether a good agreement is found and the approximate expression is able to capture both the dependency of $\Rcone$ with
$\alpha$ and $b$, as well as the order of magnitude of the optimal angle and its variation with $b$ and $L/a$.
%
% This calculation confirms that a finite slip tends to diminish the
% optimization amplitude, although this effect is quantitatively
% relevant only for very small slip lengths.

%Our findings are summarized in  Fig.~\ref{fig:comparison}.a: 
%the biconical channel has minimal hydrodynamic resistance for an optimal angle $\alopt$, which decreases in value upon increasing cone or slip lengthes. 
%Our simplified model (points) captures those main trends in a near quantitative manner. 

\section{Discussion}

Altogether, our results show that the hourglass geometry, associated with small surface friction, does optimize the 
permeability by reducing considerably the \df{magnitude} of entrance effects. A small opening angle in the range $5-20^\circ$, depending
on the precise \df{geometry and boundary conditions}, can increase the permeability by a large
factor, reaching hundreds of percents for typical parameters.

To conclude, we 
%come back to the system of inspiration, and 
discuss the relevance of these effects for the shape and the hydrodynamic permeability of aquaporins.
%hydrodynamic permeability of AQP and discuss whether these principles may
%be relevant to the real shape of aquaporins.
%Coming back to AQP, how does these principles compare to the real shape of aquaporins ?
% the biconical model can give insight into the geometry of real aquaporins?
%discuss slip length expected to be large.
%{\color{red} We have explored}
As \df{explained} in the introduction, AQPs have hourglass shapes
resembling the model geometry considered here. Furthermore, due to 
their \blue{mostly} hydrophobic 
inner surface, a small friction, and large slip length, is expected at
\blue{the cone walls}. Overall, they exhibit the main ingredients associated
with permeability optimization discussed above.
In order to push further the comparison, we have extracted some generic shape parameters of a large variety of AQPs. 
%Figure~\ref{fig:AQP} illustrates the four steps taken in building a minimal model of entrance effects within the continuum hydrodynamic description.
As illustrated in Fig. \ref{fig:AQP}, and detailed in the method section, we used 
%have extracted a 
%The starting points are 
molecular structures obtained from high-precision X-ray crystallography -- 
which are available for several aquaporins, and obtained the 
%From this wealth of information, we retain only
radius profile of the channel, as estimated by the HOLE program \cite{Smart1996}.
Though irregular, those profiles can be divided into three parts:
a central (molecular) constriction,
and two vestibules with roughly conical shapes.  
Although 
matching an atomically detailed structure to a simplified geometry involves a certain degree of arbitrariness (see Methods),
we estimated the $L$ and $\alpha$ parameters of our model for a series
of six aquaporins. 
The results for the opening angle $\alpha$ of the AQPs conical section
are displayed in Fig. \ref{fig:comparison}b. 
Two conclusions can be drawn from this figure: (i)
 the opening angle keeps rather low values, within the range $10-25 ^\circ$; (ii)
the optimal angle decreases as the cone gets longer.  
\comm{It is therefore striking that the adopted global geometry follows the expectations for the hydrodynamic optimization process discussed above
\lb{to minimize the entrance permeability}.

%While we cannot make a definite statement on the role of the hourglass shape for AQPs, these features are
%clearly reminiscent of the hydrodynamic optimization process discussed above.

Obviously, some more detailed features of the AQP geometry can not be discussed within the previous results.
%Other features of aquaporins geometry can not be explained in our
%model. 
In particular, the structure and shape selection of AQPs follows from a number of constraints and requirements, many of them from the molecular level and the subtle balances to achieve selectivity and efficient transport in the inner single-file constriction.
For instance,} aquaporins channels are not symmetric with respect to the membrane half-plane, as
the cone towards the cell exterior is apparently longer and more
divergent than the interior cone. 
Within our model, this could be explained only if the inner and outer cylinders radii were different. 
While this is often the case, 
we have considered only the average radius of the central portion, 
so as to keep a small number of parameters. 

\section{Conclusion}

The aim of this work was to determine the effect of geometry and
boundary conditions on hydrodynamic entrance effects in biconical nanochannels.  
Using finite element calculations, 
we have shown that compared to a plain cylindrical pipe, 
a biconical channel with its optimal angle can provide a spectacular increase in hydrodynamic permeability. 
A simplified model based on entrance effects and lubrication approximation rationalizes the observed behavior. 
\comm{Although speculative, this could indicate that the hourglass geometry
of aquaporins results from a shape optimization, in order to reduce end effects and maximize water permeability.}

Among transmembrane proteins, and  ionic channels in particular, 
examples abound where the particular function - ion selectivity for instance- 
is tied to a specific feature of the molecular architecture. 
Yet it remains worth wondering, as we have done here, whether generic factors such as viscous dissipation 
could be the driving force behind the shapes fine-tuned by evolution. 
Finally, in a more general biomimetic perspective, this work provides
guidelines to optimize the performances of artificial nanopores.
Hydrodynamic permeability is indeed a crucial property for
applications in water desalination, ultrafiltration or energy
harvesting from salinity gradients.
For those particularly relevant nanochannels whose inner dissipation is extremely low~\cite{Hummer2001,Falk2010},
it is of prime importance to reduce entrance effects,
because they control the overall permeability of the membrane.
% \blue{Finally, in a more general biomimetic perspective, this work provides
% guidelines to optimize the performances of artificial
% nanopores. Hydrodynamic permeability is indeed a crucial property for
% applications in water desalination, ultrafiltration or energy
% harvesting from salinity gradients. Since experiments
% \cite{Majumder2005,Holt2006} and simulations
% \cite{Hummer2001,Falk2010} indicate that some nanochannels can exhibit
% extremely low intrinsic dissipation, it is of crucial importance to
% optimize entrance effects}, \lb{ in order to optimize the overall membrane permeability}.

%%%
\begin{materials}

\section{Numerical calculations}

We study low-Reynolds numbers flows in two-dimensional axisymmetric geometry. The system of interest includes two reservoirs separated by a membrane containing one biconical nanochannel. We used finite-element method (COMSOL) to solve the Stokes equation 

\begin{equation}
\vc{\nabla} p = \eta \: \Delta \vc{v}, 
\end{equation}
where $\vc{v}$ is the fluid velocity, $p$ the pressure and $\eta$ the
fluid  viscosity. We imposed a flow rate $Q$ through the channel and measured the pressure drop across the membrane. In order to avoid finite size effects, we chose a reservoir  much larger than the pore radius. In this study we have imposed the following hydrodynamic boundary conditions:
\begin{itemize}
\item No-slip boundary condition:  This BC supposes that the fluid has zero velocity relative to the boundary, $v_w=0$ at the wall. 

\item Partial-slip boundary condition: Deviations from the no-slip
  hypothesis have been predicted theoretically and observed
  experimentally at the nanometer scale. First, assume that the
  tangential force per unit area exerted by the liquid on the solid
  surface is proportional to the slip velocity $v_w$:
  $\sigma_{xz}=\lambda v_w$, with $\lambda$ the friction coefficient,
  $z$ the normal to the surface, $x$ the direction of the
  flow. Combining this equation with the constitutive equation for a
  bulk Newtonian fluid, $\sigma_{xz}=\eta \partial_z v_x$, we obtain
  the Navier BC: $v_w=\frac{\eta}{\lambda} \partial_z v_x =
  b \partial_z v_x$ \cite{Bocquet2007}. This equation defines the slip
  length $b = \eta/\lambda$ as the ratio between the \emph{bulk}
  liquid viscosity and the \emph{interfacial} friction
  coefficient. The slip length has a simple geometric interpretation,
  as the depth inside the solid where the linear
  extrapolation of the velocity profile vanishes. 

\item Perfect-slip boundary condition: The limit of an infinite slip
  length, or equivalently a vanishing friction coefficient, is used
  when the slip length $b$ is much larger than the characteristic length(s) of the system. For simple liquids on smooth surfaces, slip lengths up to a few tens of nanometers have been experimentally measured \cite{Bocquet2007}. 
\end{itemize}

% ... In cylindrical coordinates, for a velocity $\vc{u} = u \, \vc{e}_r + v \, \vc{e}_z$, the local dissipation rate is given explicitly by 
% \begin{equation}
% \disrate = 2 \eta \left( \left( \frac{\partial u}{\partial r} \right)^2 +\left( \frac{u}{r} \right)^2 + \left( \frac{\partial v}{\partial z} \right)^2 \right) + \eta \left( \frac{\partial u}{\partial z}+\frac{\partial v}{\partial r} \right)^2. 
% \label{eq:dissip}
% \end{equation}

\begin{figure}[h]
\centering
\includegraphics[width=0.9\linewidth]{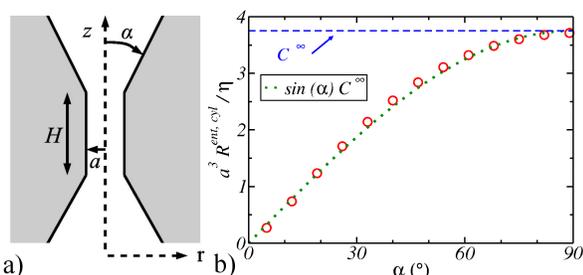}
\caption{a) Schematic of the system used to compute the
  cone-to-cylinder hydrodynamic resistance. 
b) Cone-to-cylinder resistance $\Rentcyl$ versus cone angle $\alpha$:
FE calculations (red circles) and analytical approximation by a sine
function (dotted line).} 
\label{fig:Cone_to_cylinder}
\end{figure}

\section{Cone to cylinder entrance}

In order to extract the contribution of the cone-cylinder junction to
the total resistance of the channel, we performed the following
numerical calculations. Reservoir parts are removed to eliminate the outer entrance contribution $\Rent$, leaving only a system composed of a central channel
and two truncated cones (Fig. \ref{fig:Cone_to_cylinder}-a). 
We imposed
perfect slip BC along both the cone and cylinder's walls, and
  the incoming flow fields is imposed using the far-field exact expression for frictionless cones. We varied
the angle $\alpha$ and observed that the hydrodynamic resistance of
such a junction is to a good approximation proportional to the sine
of the angle of the cone $\alpha$ (Fig. \ref{fig:Cone_to_cylinder}-b).

\section{Evaluation of aquaporins' geometrical properties}

We have chosen to divide the aquaporin in three parts; two conical
entrances and one central part (see Fig. \ref{fig:method_aqp}). Each
part of the aquaporin was linearly fitted to extract the relevant parameters.  

\begin{figure}[h]
\centering
\includegraphics[width=0.5\linewidth]{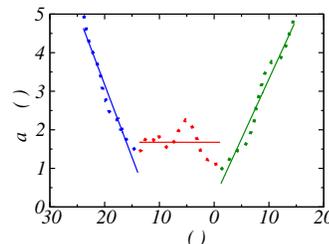}
\caption{Profile of an aquaporin (hAQP4 \cite{Cui2011}, dotted line) and linear curve fitting (solid lines).}
\label{fig:method_aqp}
\end{figure}

The central part of the aquaporin gives us the value of the central
radius $a$. From conical entrances, we extracted both the length $L$
and the angle $\alpha$. Due to the asymmetry of the aquaporin, we
obtained two values of cone length and angle for each aquaporin. 
%We are not able to explain this geometrical asymmetry with our based on Stokes' equation model. 
 
\end{materials}

\begin{acknowledgments}
This research was supported by the ERC program, project {\it Micromegas}.
We thank A. Siria and A.-L. Biance for interesting discussions. 
\end{acknowledgments}

\end{article}

\end{document}